\documentclass[prl,nobibnotes,floatfix,superscriptaddress,twocolumn]{revtex4-1}
\usepackage{amsfonts,amsmath,graphicx,epsfig,latexsym}
\usepackage{subfigure}
\usepackage{graphicx}
\usepackage{color}
\usepackage{natbib}
\usepackage{multirow}
\usepackage{latexsym,amssymb}
\usepackage{amsmath}
\usepackage{soul}
\usepackage{ulem}
\usepackage[linktocpage,bookmarksopen,bookmarksnumbered]{hyperref}
\begin{document}
\title{ Influence of magnetic ordering on the spectral properties of binary transition metal oxides}
\author{Subhasish Mandal}
\affiliation
	{Department of Physics and Astronomy, Rutgers University, Piscataway,  USA}

\author{Kristjan Haule}
\affiliation
{Department of Physics and Astronomy, Rutgers University, Piscataway,  USA}

\author{Karin M. Rabe}
\affiliation
{Department of Physics and Astronomy, Rutgers University, Piscataway,  USA}

\author{ David Vanderbilt}
\affiliation
{Department of Physics and Astronomy, Rutgers University, Piscataway,  USA}

\begin{abstract}
{\footnotesize


Using the {\it ab initio} embedded DMFT (eDMFT) approach, we study the effect of long-range magnetic ordering on the spectral properties in the binary transition metal oxides, and find that the most significant changes appear in the momentum resolved spectral functions, which sharpen into quite well-defined bands in the antiferromagnetic (AFM) phase. The strongest change across the transition is found at the topmost valence band edge (VBE), which is commonly associated with the Zhang-Rice bound state. This VBE state strengthens in the AFM phase, but only for the minority spin component, which is subject to stronger fluctuations. A similar hybridized VBE state also appears in the DFT single-particle description of the AFM phase, but gets much stronger and acquires a well-defined energy in the eDMFT description. 

}
\end{abstract}


\maketitle

\newpage


 {\it Introduction:} Understanding the effects of long-range magnetic ordering on the spectral properties of solids is important for a wide class of materials, including but not limited to the high T$_c$ cuprates and other transition-metal oxides. This has been a daunting task, as very often the magnetic transition is accompanied by a concurrent structural phase transition, with the latter often having a stronger effect. 
 One can avoid such difficulties in the binary transition-metal oxides (TMOs) such as MnO, NiO, FeO, and CoO, where the crystal structure remains almost identical during the magnetic transition. However, the quantitative description of the quasiparticle excitations in either the paramagnetic (PM) or antiferromagnetic (AFM) phase poses a theoretical challenge, since for many of the TMOs, the conventional density-functional theory (DFT) fails to predict the correct ground-state properties. Various beyond-DFT theories, such as the GW-quasiparticle approximation~\cite{PhysRevB.82.045108,PhysRevB.79.235114,PhysRevB.86.235122,PhysRevB.78.155112,PhysRevB.91.115105}, hybrid functionals~\cite{PhysRevB.92.115118,PhysRevB.74.155108} or DFT+U~\cite{Anisimov_1997,doi:10.1002/qua.24521,Gopal_2017} are now available that can correctly reproduce the photoemission and inverse photo-emission (PES/IPES) experiments for the less correlated AFM phase, and can differentiate between a charge-transfer or Mott-Hubbard nature of the gap~\cite{PhysRevB.82.045108,PhysRevB.79.235114,PhysRevB.86.235122,PhysRevB.78.155112,PhysRevB.91.115105,PhysRevB.92.115118,PhysRevB.74.155108,Anisimov_1997,doi:10.1002/qua.24521,Gopal_2017,PhysRevB.54.13566,arxiv-SM}. The problem is more acute for the PM phase, where none of these first-principles methods can properly capture the fluctuating moment in time, and fail to open the correlated charge gap, although all four TMOs remain insulators even above the Neel temperature (T$_N$). The exception is the dynamical mean-field theory (DMFT) in combination with DFT, where one does not need to know the broken-symmetry configuration {\it a-priori} and can successfully describe both the PM and AFM phases of a correlated material~\cite{haule3,PhysRevLett.119.067004,haule_spin,PhysRevB.90.060501,PhysRevB.90.060501,Kunes:2008bh,Ba122-haule}. 
 
  The strongest temperature dependence of the photoemission in insulating TMOs has been noticed for the first valence peak, which corresponds to transitions out of the valence band edge (VBE)~\cite{Kuo2017}. Theoretically, this peak is due to strong hybridization of oxygen-\textit{p} and TM 3{\it d} orbitals, and has been commonly associated with the Zhang-Rice (ZR) singlet~\cite{PhysRevB.37.3759}. The latter was introduced for hole-doped cuprates as a bound singlet state composed of an O-\textit{p} hole and a Cu-\textit{d} electron. The ZR state has been experimentally observed in several high-T$_c$ compounds~\cite{PhysRevB.37.5158,PhysRevLett.78.1126,PhysRevLett.87.237003,PhysRevB.51.8529,Chakhalian1114,ZR-Saha}. The concept of the ZR state was extended to other TMOs in Ref.~\cite{PhysRevLett.72.2600,PhysRevLett.100.066406,PhysRevLett.99.156404}, but the effect of long-range order on this state has not been discussed.

 In spite of several experimental studies of TMOs, the effect of magnetic ordering on the photoemission and inverse photoemission is still controversial. In a series of experiments, Jauch and Reehuis showed that the AFM ordering can have a profound effect on the electronic charge distribution, but only in some specific members of the late TMO family~\cite{PhysRevB.65.125111,PhysRevB.70.195121,PhysRevB.67.184420}. For CoO and MnO, they found a significant change between the PM and AFM states, while for NiO there was almost no difference in the distribution of electron density. In contrast, Shen {\it et al.} found that the AFM ordering had no significant effect on the electronic structure in CoO~\cite{PhysRevB.42.1817}. For NiO the situation is similarly controversial: Tjernberg {\it et al.}~\cite{PhysRevB.54.10245}  found no significant change due to magnetic ordering, while the recent experiment by Kuo {\it et al.}~\cite{Kuo2017} showed the development of a new peak in the valence-band spectra upon cooling, which was explained by the appearance of long-range magnetic ordering. Previous LDA+DMFT studies of TMOs ~\cite{PhysRevLett.100.066406,PhysRevLett.99.156404,PhysRevB.96.045111,PhysRevLett.108.026403,PhysRevB.74.195114,PhysRevLett.109.186401} were focused on the PM phase, and the effect of magnetic ordering on spectral properties was largely ignored
 
 In this letter, we thoroughly examine the difference between the ordered and disordered magnetic configurations by investigating the spectral functions, densities of state (DOS), and optical properties in both the PM and AFM phases of TMOs, using the DFT-embedded-DMFT (eDMFT) approach.  This is a charge self-consistent variant of DFT+DMFT that includes exact double-counting corrections between the two approaches, and is derived from the stationary eDMFT functional~\cite{eDMFT2010,eDMFT2018}.
 We find that the spectral function at the VBE changes significantly near the $\Gamma$-point in all four TMOs, while the size of the insulating gap remains unchanged across the transition. We also find that the temperature dependence of the VBE feature originates in the minority spin fluctuation of the hybridized state between O-2$p$ and TM  3\textit{d}.

{\it Method and Structural Details:} 
The Coulomb interaction $U$ and Hund's coupling $J_{H}$  are computed by the self-consistent constrained-eDMFT method (See S.I for more details about the computations), with the estimated value $U=10\,$eV and $J_H=1\,$eV in all four TMOs.
All calculations are performed at 300K on the experimental crystal structures, which are obtained from Ref.~\cite{PhysRevB.74.155108}. The lattice constants are a=4.445~$\textrm{\AA}$~\cite{MnO_expt}, 4.334~$\textrm{\AA}$~\cite{FeO_expt}, 4.254~$\textrm{\AA}$~\cite{CoO_expt}, and 4.171~$\textrm{\AA}$~\cite{NiO_expt}, for MnO, FeO, CoO, and NiO respectively. To investigate either AFM or PM state, we consider the low-temperature structure with AFM-II magnetic ordering along [111] direction~\cite{PhysRev.110.1333}, which results in the rhombohedral (R3m) symmetry, with two transition metal ions in the unit cell. 

\begin{figure*}
\includegraphics[width=460pt, angle=0]{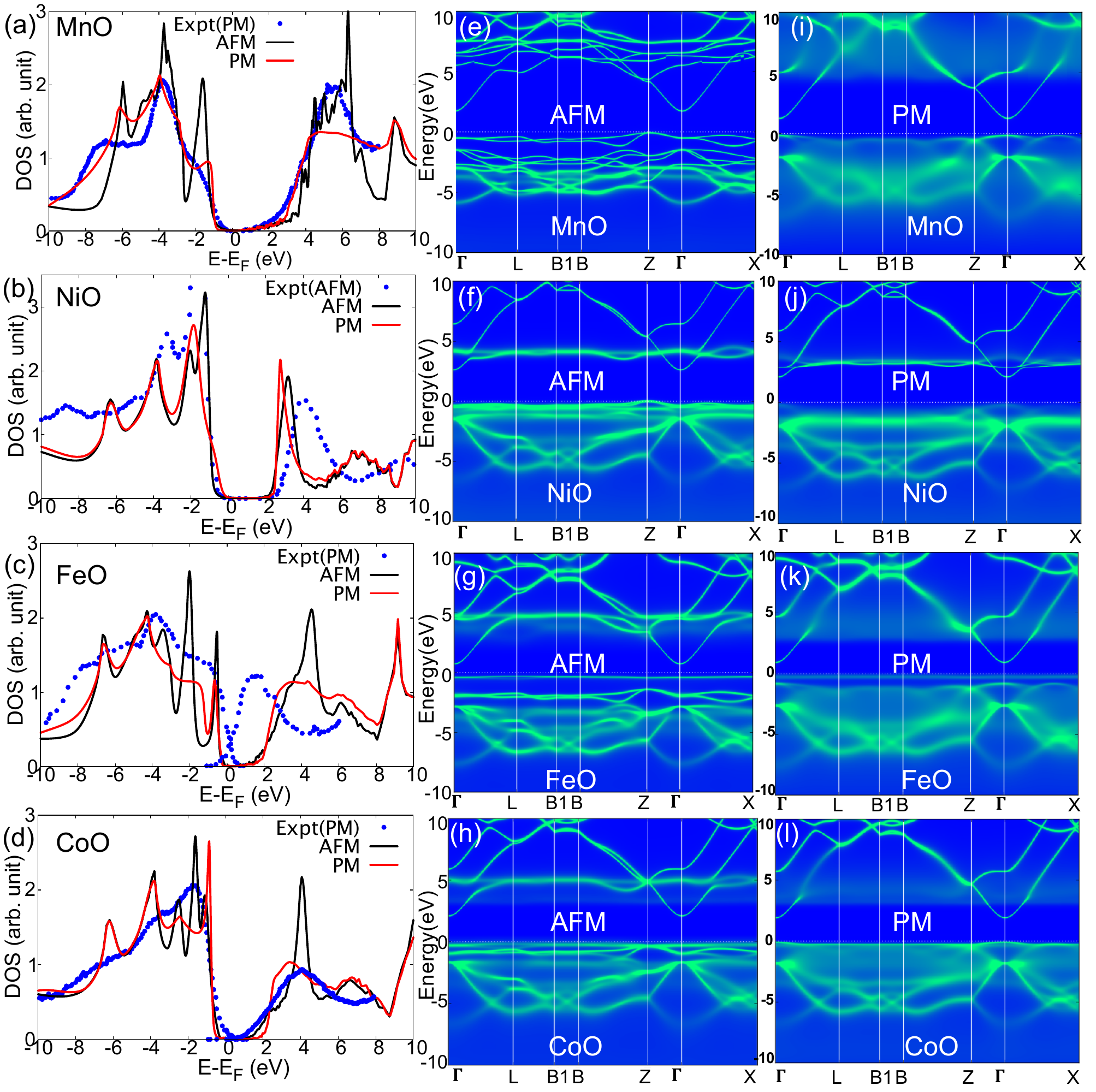}
\caption{ (Color online)
Total density of states (states/eV) as computed in 
eDMFT for (a) MnO, (b)NiO, (c) FeO, and (d)CoO. Blue dots indicate photoemission and inverse photoemission data in arbitrary unit as obtained from Ref\cite{PhysRevB.44.1530,PhysRevLett.53.2339,PhysRevB.44.6090,Zimmermann_1999} for MnO, NiO, CoO, and FeO respectively. eDMFT computed spectral functions for corresponding compounds in the antiferromagnetic (AFM) (left: e-h) and paramagnetic (PM) (right: i-l) phases. 
}
\end{figure*}


\begin{figure}
\includegraphics[width=240pt, angle=0]{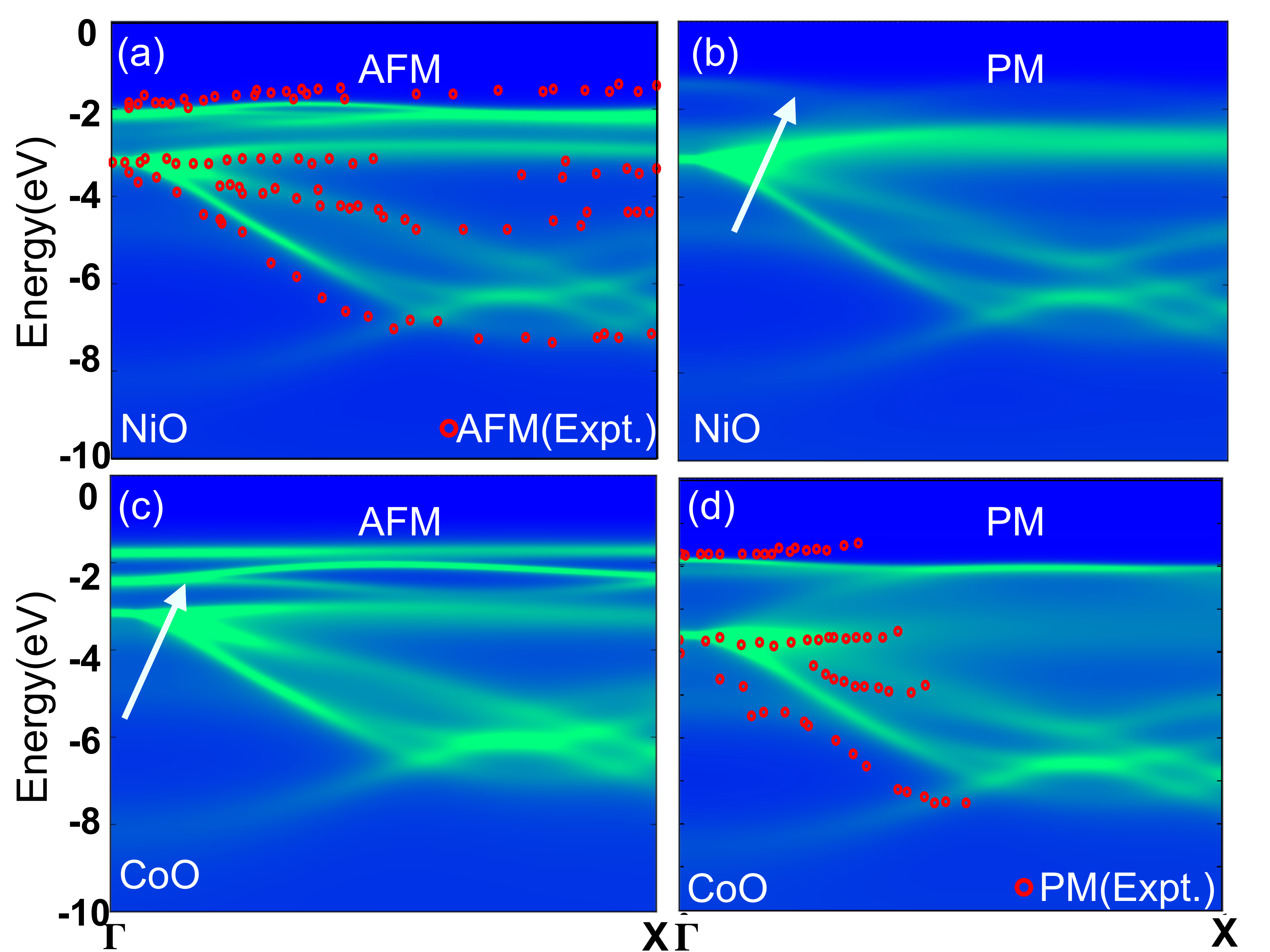}
\caption{ (Color online)
Influence of long-range magnetic ordering  in the eDMFT computed spectral function (green) of the first valence state, which is commonly associated as the Zhang-Rice state. Red dots are the experimental ARPES data by Shen {\it et al.} in NiO (up) and CoO (down) for AFM (left) and PM (right) phase respectively, which are reproduced from Ref.~\cite{PhysRevB.44.3604} and Ref.~\cite{PhysRevB.42.1817} for NiO and CoO. White arrows show the change of the VBE in two phases. 
}
\end{figure}

{\it Density of States:} In Fig. 1(a-d) we describe eDMFT computed total DOS at 300K for both PM and AFM phases of TMOs. We notice in passing that the temperature dependence of the theoretical spectra within the same phase (PM or AFM) is very weak, and the large change is seen only when we cross the phase boundary.
The DOS is compared with the experimental photoemission (PES) and inverse-photoemission (IPES), which were performed  at 300K for MnO, NiO, CoO, and FeO, and are obtained from Refs.~\cite{PhysRevB.44.1530,PhysRevLett.53.2339,PhysRevB.44.6090,Zimmermann_1999}, respectively. As the experimental PES/IPES have the arbitrary unit, we normalize them to align with the computed DOS. We label the experimental PES/IPES with their phase in which the measurement was taken (PM or AFM) and notice that the experimental T$_N$ for
MnO, FeO, CoO, and NiO are 122K, 192K, 290K, and 523K, respectively. Therefore, except for NiO, all PES/IPES at room temperature were performed in the PM phase.

As we did not account for the matrix element effect in PES/IPES processes, we can not expect a very precise match between the peak intensities of the theoretical DOS and PES/IPES, but we rather concentrate on the peak positions and overall weight distribution.

Overall, eDMFT shows a very good agreement with the PES/IPES peak positions (Fig. 1a-d), except for FeO. 
This is likely 
because FeO crystals tend to be non-stoichiometric~\cite{FeO_nonstochiometric}, hence their IPES spectra is likely shifted down for approximately $2\,$eV. Such a simple shift of IPES would greatly improve the agreement between the theory and experiment.
A detail head-to-head comparisons with experimental PES/IPES and the computed DOS in the AFM phase of the TMOs are described in a separate paper~\cite{arxiv-SM}.

From Fig. 1(a-d) we find that theoretically, the insulating charge gap for all TMOs remains almost unchanged across the PM to AFM transition. However, a few differences in the local spectra are notices: i) in NiO, and CoO the first peak in the valence band, i.e., VBE splits into two when going from PM to AFM phase, while in MnO and FeO the same peak just sharpens and strengthens in the AFM phase, 
ii) the first unoccupied state appears as a very sharp band in momentum resolved spectra in either phase (see Fig. 1e-l), while it appears as a small shoulder, weakly increasing with increasing energy. Our analysis shows that it is of TM $4s$ character, and is weakly temperature-dependent. iii) In contrast, the next unoccupied state (peak around 4eV in FeO, CoO, and 3eV or 5eV in NiO or MnO) is mostly of TM 3$d$ character, and it considerably sharpens in AFM phase in all TMOs except for NiO.

We notice that DFT+DMFT with popular fully localized double-counting scheme~\cite{FLL} compares poorly with experiment (Fig. S1). Although various alternative approaches, such as the self-interaction corrected DFT+DMFT\cite{sicDMFT} or GW+DMFT \cite{PhysRevB.93.235138} were recently shown to improve the spectra, we show here that the {\it exact}-double-counting~\cite{exact} gives similar agreement with experiment, and places the VBE properly, and at the same time gives a sizable insulating gap, which compares well with the experimental PES/IPES.

\begin{figure}
\includegraphics[width=240pt, angle=0]{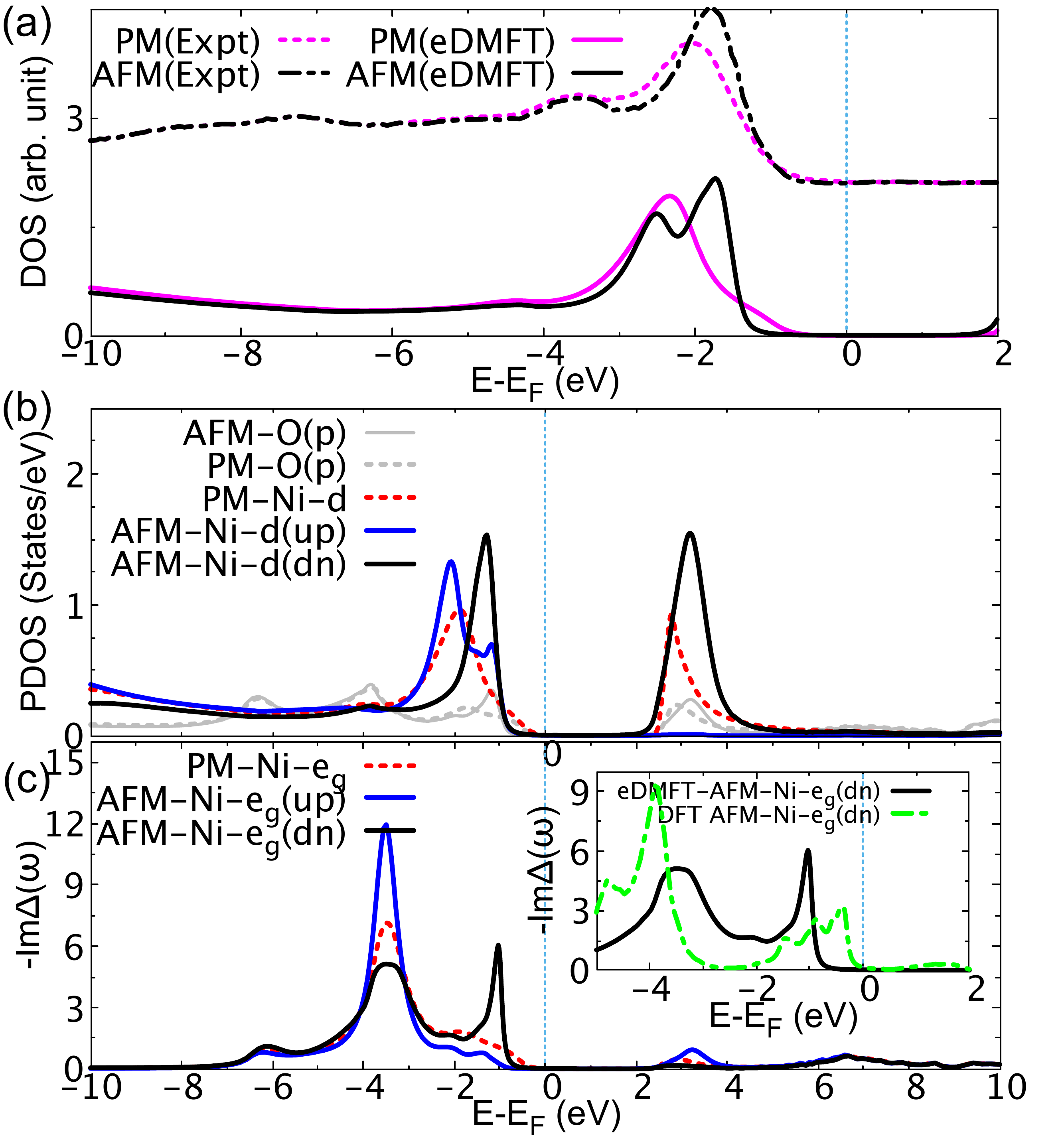}
\caption{ (Color online)
 (a) eDMFT computed total 3-{\it d} density of states (DOS) per atom and comparison with experiment for NiO. The experimental data are obtained from Ref.~\cite{Kuo2017}. (b) Computed partial DOS of $O$-2$p$ (in gray) and 3{\it d} orbitals  and (c) eDMFT hybridization function for the majority (up) and minority (down) components of spin for $e_g$ orbital shown in solid blue and black lines respectively in NiO for the AFM phase; same quantities in the PM phase are shown in red dotted lines. The inset shows zoomed-in hybridization function as computed with DFT (green) and eDMFT (black) for minority spin in Ni-e$_g$. Blue dotted lines indicate the Fermi level(E$_F$). The E$_F$ in DFT plot is shifted within the gap for clarity.
}
\end{figure}

{\it Spectral function:} Fig. 1e-l show momentum resolved spectral functions in the AFM (left) and PM (right) phase for all four TMOs.
The spectral functions in the PM phase are more diffusive or incoherent as compared to AFM phase because the fluctuating magnetic moment can not be described in terms of Bloch bands, while the AFM phase is much more mean-field like, and is well described in the band picture. 
We notice that the spectral function of MnO in AFM phase is much sharper than the others, because the entire fluctuating moment of $5/2$ orders in MnO, hence the system is orbitally a singlet, which makes the system more mean-field-like and less correlated.
We also notice that the most significant change in the peak dispersion is along Z-$\Gamma$-X direction, where the VBE has a maximum at the $\Gamma$ point in the PM phase, while it has a minimum in AFM phase of MnO and NiO. In FeO and CoO, the VBE becomes essentially momentum independent in the AFM phase, hence local to the single unit cell. This flat band gives rise to a sharp peak in the occupied DOS near E$_F$. Such a flat band is also observed in GW calculation done on top of hybrid functional~\cite{PhysRevB.86.235122}, however not found by DFT+U method. 
It is noteworthy to mention that we do not find the spectral weight to vanish near the $\Gamma$ point in either PM or AFM phase, which was previously predicted for some cuprates~\cite{PhysRevLett.100.066406}, but later assigned to the matrix element effects~\cite{PhysRevLett.113.137001,PhysRevLett.99.237002}.

{\it Comparison with experiments:} Next, we compare eDMFT computed spectral functions for both AFM and PM phases with experimental ARPES spectra, which were obtained at room temperature for NiO and CoO ~\cite{PhysRevB.44.3604, PhysRevB.42.1817}, the only two (out of four) compounds for which the  ARPES data are available in the literature.
As the position of the chemical potential in the insulating gap at low-T is arbitrary, we vertically align the theoretical spectra to best match with ARPES spectra at the $\Gamma$ point.
In both NiO and CoO, we notice two types of states - at the top is the narrow VBE band of mostly $3d$-character, and deeper below E$_F$ are several more dispersing bands of mostly O-$2p$ character. 

In NiO, VBE is very weak in the PM phase and dispersing downward from $\Gamma$ to $X$ (white arrow in Fig. 2b), while it is dispersing upward around $\Gamma$ in the AFM phase, and is much stronger. The experimental ARPES spectra were taken in the AFM phase and clearly matches much better with theoretical AFM calculation, as the VBE state is clearly resolved and is dispersing up from $\Gamma$ point, similar to the computed eDMFT spectra.  An additional extremely weak spectrum was observed for the uppermost valence band (not shown here), which was only noticed for selected photon energies and certain emission angles in the ARPES experiment~\cite{PhysRevLett.99.156404,PhysRevB.44.3604}. The uncertainties of this spectrum were discussed in Ref.~\cite{PhysRevLett.99.156404,PhysRevB.44.3604}.

For CoO (Fig. 2c-d), theory predicts that VBE splits into two peaks in the AFM phase (see white arrow), which has not been detected in ARPES, as the experiment is performed above the Neel temperature. The agreement with experiment is good only when we consider the PM state, as compatible with measurement. As a consequence, the eDMFT theory predicts that the ARPES experiment in the AFM phase should see an additional flat band between VBE and oxygen $2p$ bands.

While NiO ARPES data are available only in the AFM phase, the integrated PES was recently measured in both PM and AFM phase~\cite{Kuo2017}, and a profound enhancement of the VBE peak intensity was observed upon cooling through the phase transition.
To enhance this difference across the transition, we plot in Fig. 3a the partial $3d$-DOS in both phases, and compare it with the PES from Ref.~\cite{Kuo2017}.
It is clear from Fig. 3a that the intensity of VBE considerably sharpens in the AFM phase in both experiment and theory. The splitting of the VBE is however not observed in the experiment.

{\it Hybridization: }To gain further insights into the VBE splitting in the AFM phase of NiO, we resolve the partial DOS into spin-majority (up) and spin-minority (dn) contribution in Fig. 3b. We notice that the first VBE peaks come from spin-minority (in black), and the second from the majority component (in blue). The latter is centered close to the PM peak (in red), while the minority peak is the first excitation in the AFM phase. We notice that VBE peak has a substantial admixture of oxygen-$p$ DOS, hence it is a hybrid of the 3$d$ and O-2$p$ orbital.
To better understand the role of hybridization between oxygen-$p$ and $3d$ orbitals, we display in Fig. 3c (Fig. S2)  the eDMFT hybridization function in both PM and AFM phase for NiO (all four TMOs). We notice that hybridization at the VBE is relatively small in the PM phase as well as in the AFM phase for the spin-majority channel. This could be explained by the difficulty of screening a large magnetic moment by itinerant states, such as oxygen $p$-bands, an effect noticed early on by R. Schrieffer~\cite{doi:10.1063/1.1709517} in the context of Kondo effect. However, quantum fluctuations of the spin-minority states are usually larger than that for more mean-field like majority component. Therefore, it is perhaps not surprising that the spin-minority hybridization develops a very sharp peak at the energy of the VBE. This proves that the VBE is due to very strong hybridization between TM 3$d$ and 2$p$ oxygen orbitals, however, its origin does not need to be many-body in nature, such as the Zhang-Rice type singlet/bound state. In the inset, we describe the same hybridization function calculated on the DFT solution, which also shows a prominent peak near the VBE. Hence at least part of this strong hybridization is due to enhanced hopping described in the single-particle theories, however, the many-body effects contained in DMFT do sharpen the VBE peak,  and make more well defined in energy. At the same time, the momentum resolved spectral function shows that the momentum dependence of the VBE state is much weaker, therefore, this state becomes more local to a single unit cell. So we can conclude that the many-body effects make this $3d$-$2p$ hybridized state longer-lived and more localized within the single unit cell.

{\it Optics:} Optical absorption measurements are easier to perform on such large gap insulators than ARPES measurements, it is therefore interesting to check if the effects of long-range order can be seen by optics. We calculated optical absorption (Fig. S3) and notice that the difference between the PM and AFM state is very small, and hence the effects of the long-range order would be hard to find by optical experiments.

{\it Conclusions:}
 In conclusion, we find the most significant changes across the magnetic phase transition in TMOs are in the momentum resolved spectral functions, while these changes are insignificant in optical absorption, as the size of the charge gap remains unchanged. The spectral function is very incoherent in the PM phase, but sharpens into quite well-defined bands in the AFM phase. The strongest change across the transition is found at the VBE, which is commonly associated with the Zhang-Rice singlet state, as it comes from strong hybridization between oxygen-\textit{p} and transition metal 3\textit{d} orbitals. This VBE state appears as a relatively weak photoemission peak in the PM phase, but strengthens in the AFM phase only in the minority spin channel, which is subject to stronger fluctuations. We point out that similar hybridized VBE state also appear in the DFT single-particle description of the AFM phase. Hence, its origin is not purely many-body in nature. However, in the eDMFT description, this state acquires a stronger intensity, a well-defined energy, and an extremely flat momentum dispersion.

{\it Note -} While preparing this manuscript we became aware of a recent arXiv article~\cite{NiO_arxiv} in which  a similar VBE peak splitting in the AFM phase of NiO is described using a GW+DMFT approach.   

\section{Acknowledgment } 
We thank G. L. Pascut for helpful discussions. The computations were performed at the XSEDE, Rutgers HPC (RUPC). This research also used resources from the Rutgers Discovery Informatics Institute\cite{RDI2}, which are supported by Rutgers and the State of New Jersey.  This research was funded by NSF DMREF DMR-1629059 and NSF DMREF DMR-1629346.

\bibliography{SM-bib}
\section{Supplementary Information }   

{\it Method and Structural Details:} 
In DFT+eDMFT method~\cite{eDMFT2010,eDMFT2018} we use LDA functional, and the LAPW basis set as implemented in WIEN2k~\cite{WIEN2k}. The continuous time quantum Monte Carlo method~\cite{haule2007} is used to solve the quantum impurity problem that is embedded within the Dyson equation for the solid, to obtain the local self-energy for the TM $d$ orbitals. The self-energy is then analytically continued with maximum entropy method from the imaginary to the real axis, continuing the local cumulant function, to obtain the partial density of states.  A fine k-point mesh of at least 10$\times$ 10 $\times$ 10 k-points in Monkhorst-Pack k-point grid and a total 100 million Monte Carlo steps for each iteration are used for the AFM phase of the TMO at T=300K.The Coulomb interaction $U$ and Hund's coupling $J_{H}$  are computed by the self-consistent constrained-eDMFT method, with the estimated value $U=10\,$eV and $J_H=1\,$eV in all four TMOs. All calculations are performed at 300K on the experimental crystal structures, which are obtained from Ref.~\cite{PhysRevB.74.155108}. The lattice constants are a=4.445~$\textrm{\AA}$~\cite{MnO_expt}, 4.334~$\textrm{\AA}$~\cite{FeO_expt}, 4.254~$\textrm{\AA}$~\cite{CoO_expt}, and 4.171~$\textrm{\AA}$~\cite{NiO_expt}, for MnO, FeO, CoO, and NiO respectively. To investigate either AFM and PM state, we consider the low-temperature structure with AFM-II magnetic ordering along [111] direction~\cite{PhysRev.110.1333}, which results in the rhombohedral (R3m) symmetry, with two transition metal ions in the unit cell.  \\

{\it DOS:} In Fig. S1 we describe comparison of DOS with FLL and exact double counting for NiO in eDMFT for AFM phase. In Fig. S2 we describe the PDOS for Oxygen-2$p$ and TM 3$d$ for both PM and AFM phases. We also describe  hybridization function for the majority/up (minority/down) components of spin in the AFM phase in solid blue (black) lines respectively for (a-c)MnO, (d-f)NiO, (g-i)FeO, and (j-l)CoO; similar quantities for the PM phase are also shown for $e_g$ (red) and t$_{2g}$ orbital (pink). 
\\

{\it Optics: }
To compute the absorption coefficient within eDMFT, we obtain the imaginary part of the dielectric function from the real part for the optical conductivity and then perform the Kramers-Kronig (KK) operations. Fig S3 describes the the absorption coefficient within eDMFT for both PM and AFM phase. For NiO and CoO, the experimental absorption coefficients are extracted from Powell~\textit{et.al.}~\cite{PhysRevB.2.2182}, which were obtained from the measured reflectively spectra. For MnO, the optical spectra is extracted from figure in R\"old~\textit{et al.}~\cite{PhysRevB.86.235122}, where the measurements by Ksendzov~\textit{et al.}~\cite{MnO-optics} were reproduced. The original data for MnO is not currently accessible and reliable data for stoichiometric FeO is not available.

\begin{figure}
\includegraphics[width=220pt, angle=0]{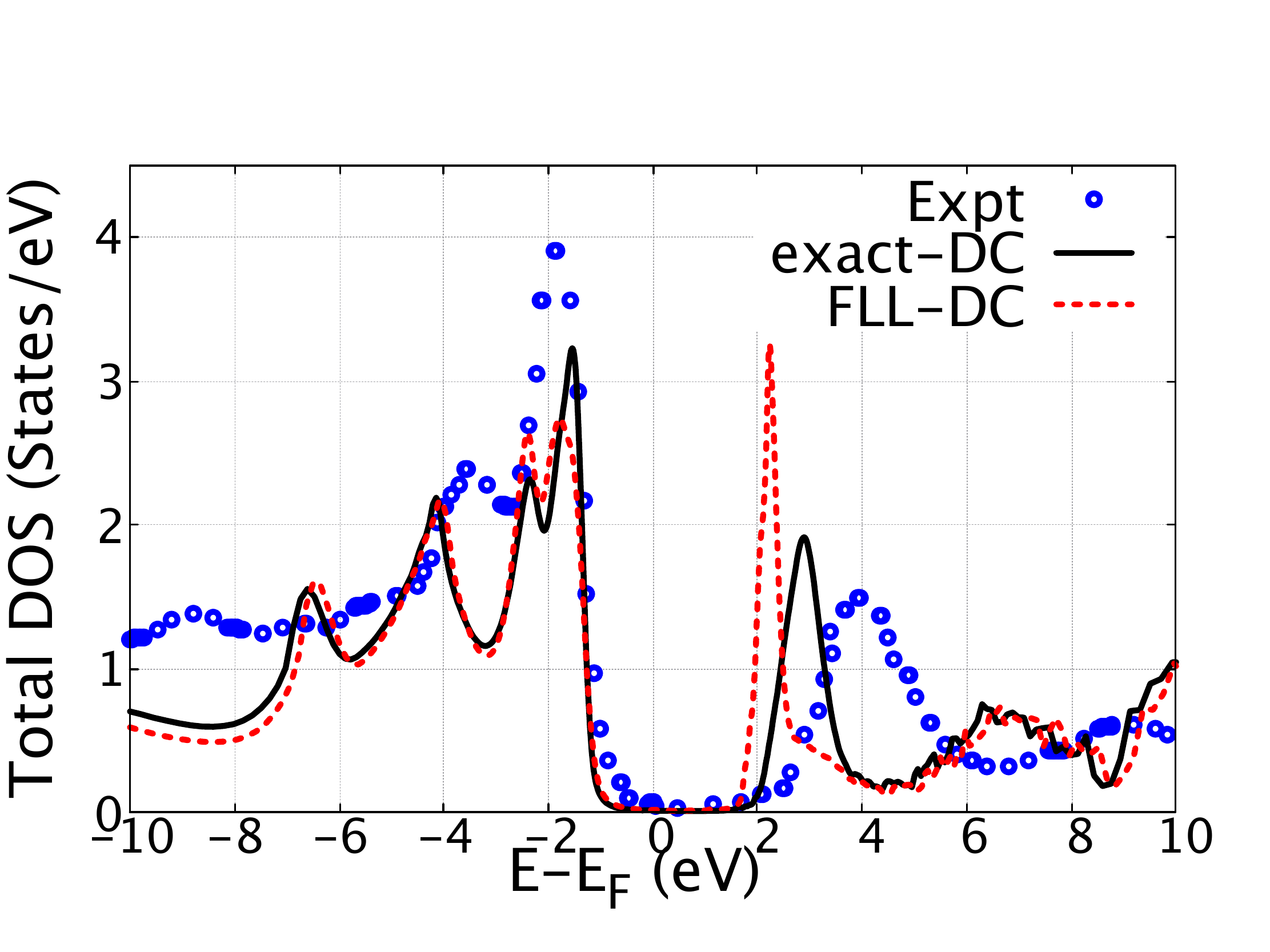}
\caption{ (Color online)
Comparison with FLL and exact double counting for NiO in eDMFT for AFM phase.}
\end{figure}

\begin{figure*}
\includegraphics[width=460pt, angle=0]{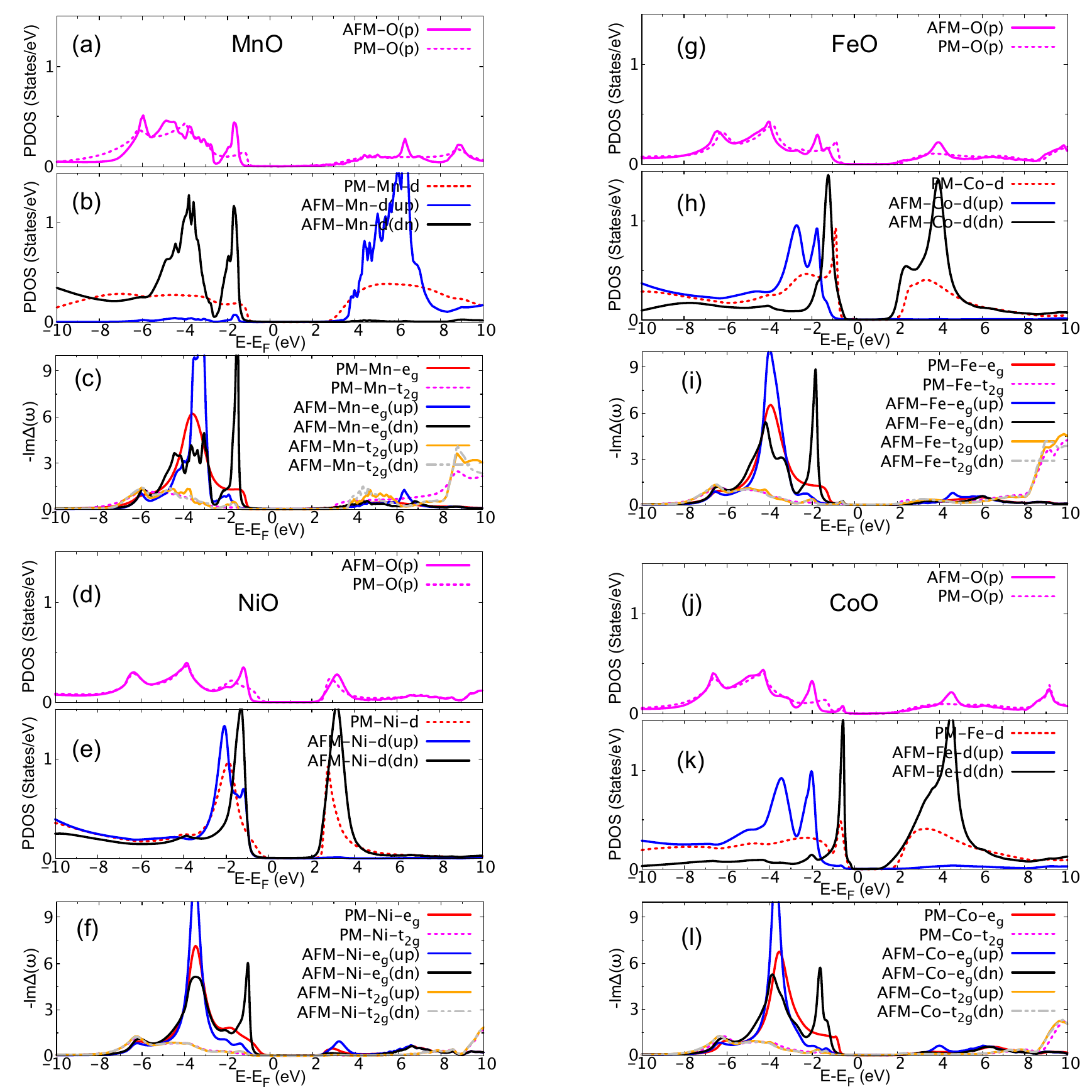}
\caption{ (Color online)  eDMFT computed partial density of states (PDOS) per atom (top two panels) and hybridization function for the majority or up (minority or down) components of spin in the AFM phase are shown in solid blue (black) lines respectively for (a-c)MnO, (d-f)NiO, (g-h)FeO, and (i-l)CoO. Similar quantities for the PM phase are also shown for $e_g$ (red) and t$_{2g}$ (pink) orbitals.
}

\end{figure*}

\begin{figure*}
\includegraphics[width=480pt, angle=0]{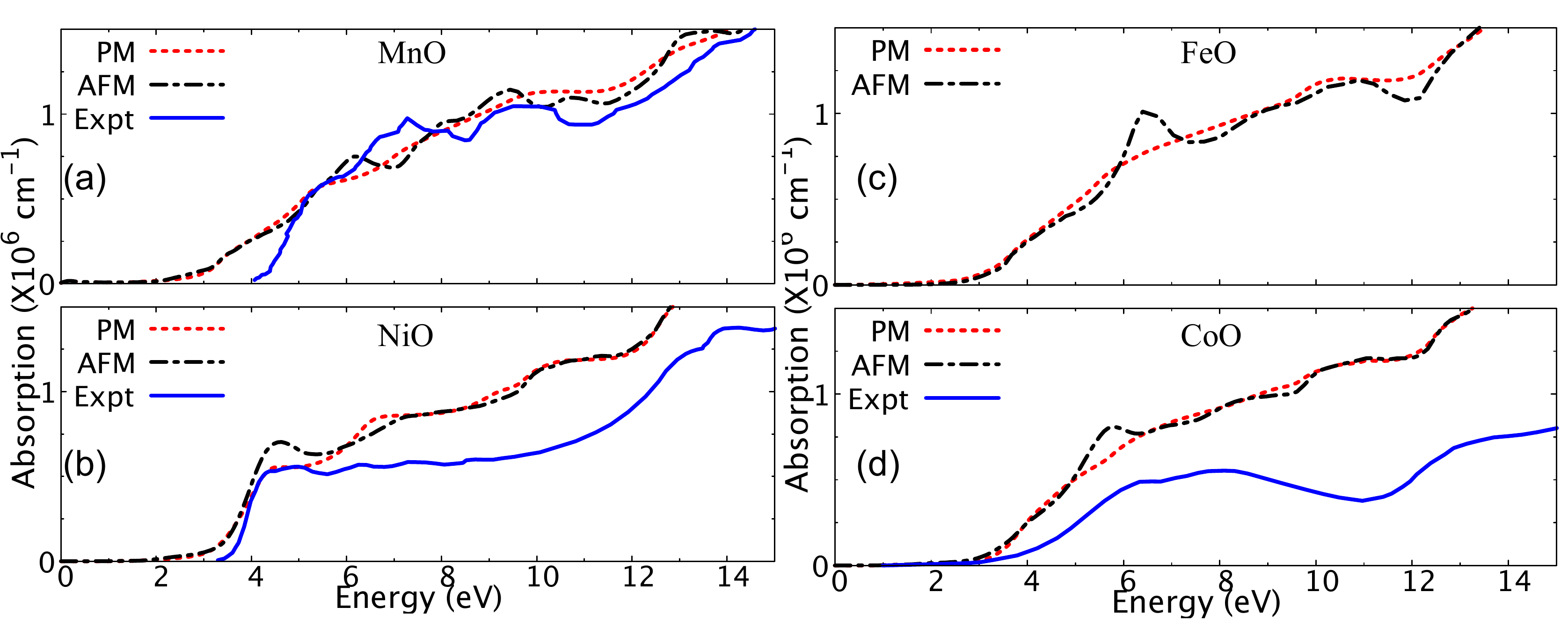}
\caption{ (Color online)
Comparison of eDMFT computed optical absorption coefficients  for all four TMOs with available experiments (in blue solid lines). Experimental data are directly obtained from Powell {\it et al.}\cite{PhysRevB.2.2182} for CoO and NiO and from  R\"odl {\it et al.}\cite{PhysRevB.86.235122} for MnO, which was used to compare with the experimental reflectivity by Ksendzov {\it et al.}\cite{MnO-optics}. 
}
\end{figure*}



\bibliography{SM-bib}

\end{document}